\documentclass[prd,preprint,amsmath,nofootinbib,superscriptaddress]{revtex4-1}
\pdfoutput=1
\usepackage{bm}
\usepackage{slashed}
\usepackage{amsmath,latexsym}

\usepackage{graphicx}
\usepackage{bbold}
\usepackage{hyperref}
\usepackage[usenames,dvipsnames]{color}

\def\OMIT#1{}

\newcommand{\bea}{\begin{eqnarray}}
\newcommand{\eea}{\end{eqnarray}}

\newcommand{\beq}{\begin{equation}}
\newcommand{\eeq}{\end{equation}}
\newcommand{\mo}{\mathcal{O}}
\newcommand{\nn}{\nonumber} 

\newcommand{\vacproj}{|0\rangle \langle 0 |}

\begin{document}

\title{Constraints on Galactic Wino Densities from Gamma Ray Lines}

\author{Matthew Baumgart, Ira Z. Rothstein, and Varun Vaidya\vspace{0.4cm}}
\affiliation{Department of Physics, Carnegie Mellon University,
    Pittsburgh, PA 15213}


\begin{abstract}
We systematically compute the annihilation rate for neutral winos into the final state $\gamma + X$, including all leading radiative corrections.  This includes both the Sommerfeld enhancement (in the decoupling limit for the Higgsino) and the resummation of the leading electroweak double logarithms. Adopting an analysis of the HESS experiment, we place constraints on the mass as a function of the wino fraction of the dark matter and the shape of the dark matter profile.  We also determine how much coring is needed in the dark matter halo to make the wino a viable candidate as a function of its mass.  Additionally, as part of our effective field theory formalism, we show that in the pure-Standard Model sector of our theory, emissions of soft Higgses are power-suppressed and that collinear Higgs emission does not contribute to leading double logs.
\end{abstract}

\maketitle

\section{Introduction}

One of the biggest challenges in contemporary high energy physics is determining the identity of dark matter (DM).  Unfortunately, all of the astrophysical and cosmological evidence for dark matter does not answer our most basic phenomenological question: does it have non-gravitational interactions with the Standard Model (SM)?   Having a new field coupled to the SM non-gravitationally offers the possibility of both a near-term discovery and the elucidation of other open questions, such as the hierarchy and strong-CP problems.  

The WIMP Miracle presents a particularly compelling link between the weak scale and dark matter (see \cite{Dimopoulos:1990gf}).  Demanding the correct relic abundance from  cosmological freeze-out leads one to an $\mo$(TeV)-mass particle with electroweak-strength coupling.  We consider here the case of one particular DM candidate, the wino, that belongs to a supersymmetric explanation of the weak scale.  Despite the lack of direct evidence for the MSSM, the discovery of a SM-like Higgs at 125 GeV leaves open the possibility of a modestly-tuned supersymmetry scenario that retains a simple mechanism of SUSY-breaking and a standard, thermal-relic DM particle \cite{minisplit}.  In fact, the only feature of the MSSM we use is the presence of a stable, electroweak triplet fermion.  Thus, our result does not depend on the larger supersymmetric story, but holds for any DM scenario with the same quantum numbers, annihilating primarily through its gauge interactions.  Extensions to scalar triplet or other SU(2) representations are straightforward.    

A nearly pure wino offers one of the simplest supersymmetric DM candidates.  In theories of anomaly-mediated SUSY-breaking, it emerges as the lightest supersymmetric particle (LSP) \cite{amsb}.  Furthermore, thermal-relic bino dark matter generically overcloses the universe by several orders of magnitude. If we assume that the wino constitutes all of
the dark matter - which is an assumption we will relax below -  and that  its relic density was set at freeze-out, then the mass is constrained to the  window $M_{\rm Wino} \equiv (M_{\chi})$ = 2.7-2.9 TeV \cite{Fan:2013faa,constraint}.

In principle, we may further constrain the wino (DM) via direct detection.  However, the cross section for  a TeV wino to   scatter off nucleons  is  $\sigma \sim 10^{-47}$ cm$^{2}$, putting it far below current limits \cite{Hisano:2011cs,Hill:2013hoa}.  However, since the wino can annihilate directly to photons, by searching for monochromatic, $\mo$(TeV) photon lines, we can hope to discover it via indirect detection.  The authors of \cite{Fan:2013faa,Cohen:2013ama} used limits from the HESS Cherenkov telescope to argue that the nonobservation of such a photon feature put wino DM in severe tension with experiment.\footnote{They also consider the constraints from continuum gamma ray emission provided by the Fermi experiment.  These are most useful for constraining low-mass, few-hundred GeV winos.  As the radiative corrections we investigate are much weaker in this regime, we do not investigate it here.}  In particular \cite{Cohen:2013ama} calculated the annihilation rate to be $\sim15\times$ larger at  $M_{\chi}$ = 3 TeV than the HESS limit.

However, as with any indirect detection experiment, we must take into account astrophysical uncertainties.  In particular, \cite{Fan:2013faa,Cohen:2013ama} consider variations of the DM halo profile for the galaxy.  While cuspy profiles are preferred by simulations, it is possible that the dark matter density flattens out to a ``core'' about the galactic center, and such a distribution would lead to fewer DM annihilation events along our line of sight to the galactic center.  To alleviate the tension with HESS, some amount of coring will be necessary to return pure wino dark matter to viability, but the question, which we answer in Section \ref{sec:conc}, is how much?

The annihilation rate of two heavy WIMPs ($M_{\chi} \gg M_W$), which are nonrelativistic, cannot be reliably calculated at tree level since it is plagued by  infrared (IR) divergences which are cut-off  by the gauge boson mass, $M_W \sim$ 100 GeV. These divergences manifest themselves as  large radiative corrections of two types.\footnote{The term "divergences" is used despite the fact that the rate is physical.}  One set comes from the potential interactions of the slowly-moving DM and scales as powers of $\frac{\alpha_W M_\chi}{M_W} \gtrsim 1$; the resummation of these  corrections results in a Sommerfeld enhancement to the rate \cite{Hisano:2004ds}.   This effect can increase the rate by as much as $\mo(10^4)$ relative to a perturbative calculation and is therefore a crucial step in analyzing wino DM.  The second type of IR sensitivity is a Sudakov double-log, $\alpha_W \log (\frac{M_\chi^2}{M_W^2})^2$, that can enter inclusive observables due to the non-singlet nature of our external states.  This effect is known as ``Bloch-Nordsieck Theorem Violation'' and is generically found in the Higgs phase  of non-Abelian gauge theories \cite{Ciafoloni, Manohar}, such as the electroweak sector.  Computations of fixed, NLO corrections to the exclusive, two-body annihilation rate found a 75\% reduction relative to tree level plus Sommerfeld enhancement \cite{Hryczuk:2011vi,Cohen:2013ama}.  This opened up the possibility that the wino could still be viable, even with a non-cored dark matter profile.  This result motivated the need for a systematic approach to the calculation of the rate since such a large radiative
correction gives the appearance of series which is diverging too soon for an asymptotic expansion with such a small coupling. 

In a previous paper, we derived a factorization theorem that was used to derive the  leading-log (LL)  semi-inclusive wino annihilation rate ($\chi^0 \chi^0 \rightarrow \gamma + X$) \cite{Baumgart:2014vma} in terms of a model-dependent set of matrix elements.  The semi-inclusive rate is  the relevant observable for constraining the wino with HESS since only a single hard photon from annihilation is measured and the resolution of the experiment ($\sim$ 400 GeV at 3 TeV) is too poor to distinguish two-body from $n$-body annihilation ({\it e.g.}~$\chi^0 \chi^0 \rightarrow \gamma + W^+ W^-$) \cite{hess}.  Despite the fact that both the Sommerfeld and Sudakov effects arise from the same hierarchy, $M_\chi \gg M_W$, they can be factorized through a mode decomposition of the relevant fields.  Physically, we are separating the regime of the slowly-moving WIMPs evolving in the presence of the electroweak potential from the highly-energetic gauge bosons in the final state.  We therefore employ a theory that combines a nonrelativistic treatment of the WIMPs, similar to NRQCD \cite{NRQCD}, with Soft-Collinear Effective Theory (SCET) \cite{SCET} for the light annihilation products.  Hybrid theories of this form have appeared in the analysis of the photon spectrum in radiative decays of quarkonium \cite{Leibovich} and electroweak SCET was elucidated in \cite{aneesh}.  More recently, other groups performed the Sudakov resummation for WIMP annihilation, albeit for the exclusive two-body rate, employing the same effective field theory (EFT) \cite{Bauer:2014ula,Ovanesyan:2014fwa}.

In this paper, we give more details of of the factorization theorem presented in \cite{Baumgart:2014vma} in Section \ref{sec:fact} and elucidate the role of the soft and collinear Higgs fields at higher orders in Section \ref{sec:higgs}.  In Section \ref{sec:ad} we  review the anomalous dimension calculations that allow Sudakov log resummation and provide additional detail on our use of the rapidity renormalization group \cite{RRG}.   To obtain a full calculation of the annihilation rate, in Section \ref{sec:se} we present our analysis of the Sommerfeld enhancement for the particular case of wino dark matter with a parametrically large Higgsino mass.  We find that compared to tree level plus Sommerfeld corrected rate, the  leading-log radiative corrections lead to a modest, few-percent reduction in the semi-inclusive rate to an observable photon.  
In Section \ref{sec:conc}, we also present exclusion plots for wino dark matter as a function of the mass and the amount of coring in the dark matter profile and the wino fraction of the dark matter.
After concluding, we detail in an appendix the derivation of the quantum mechanical potential used in our Sommerfeld enhancement calculation from the underlying, relativistic quantum field theory.

\section{Factorization}
\label{sec:fact}

To develop a factorization theorem it helps to work in an EFT, which in our case is a hybrid of SCET and NRQCD
that power counts in a double expansion in $v$, the relative velocity of the  DM particles  and $\lambda=M_W/M_\chi$.
We will work to leading order in both these parameters, which seems quite reasonable until
a discovery is made.

\subsection{NR``QCD"}

The nonrelativistic (NR) piece of the Hilbert space is described by an effective theory which is analogous to
NRQCD, but differs in two important aspects.  First off, our theory of interest is in the weakly coupled Higgs phase
and moreover, the potential between the DM particles is screened by the gauge boson mass. This latter distinction
changes the way in which we power count as we shall see below. For various power counting schemes in NR theories see 
\cite{manohar}. Despite these distinctions, the modal analysis of the effective theory follows from the general discussion in \cite{lmr} and
furthermore we will still use the acronym NRQCD.

NRQCD can be formulated in terms of three distinct types of modes,  each with a unique scaling of momenta: Potentials $(E\sim mv^2, p\sim mv)$, soft ($E\sim mv, p \sim mv)$
and ultrasoft (US) ($E\sim mv^2,p\sim mv^2$), where $v$ is the relative velocity of the massive nonrelativistic states. The massive states have energy-momenta scaling as potential modes but nonetheless are on-shell, due to their NR dispersion relation. On the other hand, the potential  gauge bosons are off-shell and can be integrated out to form non-local potentials. The soft modes are necessary to generate the correct running of the potentials, but will not play a role in the theory at hand as all of their effects
will be of higher order. The same can be said for the US modes. However, the US modes will play a crucial role in determining the gauge invariant structures
allowed in the theory.

It is important to understand that $v$ is NOT necessarily the incoming relative velocity of the NR particles. $v$ can also be the virialized velocity.
That is, if $v$ is sufficiently small then the particles can inspiral, gaining kinetic energy until the system virializes such that $V(r)\sim mv^2$.
In the Coulomb phase (or for sufficiently large source masses in the confining phase of an asymptotically free theory) we have the
relation
\beq
\frac{g^2}{r} \sim mv^2
\eeq
which leads to the scaling $v \sim \alpha$ given the Bohr radius  $r\sim 1/(\alpha m)$. If the incoming relative  velocity is large
enough then the system may not virialize. The condition for virialization is that the leading order potential be non-perturbative. In the Coulombic case this condition is
$\frac{\alpha}{v}\sim 1$.  When this condition is met we must sum the box graphs. In such cases the incoming relative
velocity becomes irrelevant for the power counting.\footnote{This is true only if we are interested in time averaged quantities. If we wanted to
track the explicit time dependence of the power counting parameter, the use of  the in-in formalism \cite{galley}
is called for.}

The NRQCD-like theory in our case is more complicated since gauge boson exchange flips a neutralino to a chargino
which is taken to be a few hundred MeV heavier. Moreover the charge state admits Coulomb exchange, although the off-shell nature
of the chargino intermediate plays the role of an IR cut-off.  Since we are only interested in working at leading order in the $v$, the
exact details of the correct effective theory will be irrelevant. All that matters for the present analysis is that there exist some
virialized velocity, $v$, which will play the role of the power counting parameter. Given that additional rungs on ladder diagrams bring inverse powers of $v$, we need to sum over an infinite ladder of potential mode gauge boson exchanges between the massive fermions resulting in the so-called Sommerfeld enhancement which we will discuss in Section \ref{sec:se}.

NRQCD is invariant under both soft and US gauge transformations.\footnote{The theory developed in \cite{lmr} is not explicitly soft
gauge invariant at the level of the action. However, the theory can be re-derived in such a way at to preserve soft gauge invariance \cite{RSS}} 
Under a US gauge transformation all fields which are left on shell can transform non-trivially.  Given that US transformations shift momenta and energy by an amount of order $mv^2$,  the potential (matter) 
fields transform under US but not soft gauge transformation. This fact will play an important role below when we write down a factorization theorem.

\subsection{SCET-II}
The momentum modes we need in our EFT are those which describe our external states and capture the dominant IR physics. The KLN theorem tells us that the IR physics is described by ``dangerous states" \cite{Weinberg}, which in our case are modes for which, either all components of momenta are small (soft modes)  or modes whose momentum is collinear with the observed photon.  What we mean by small is decided by the power counting parameter $\lambda(\equiv M_W/M_\chi)$.  For SCET, we choose to choose a lightlike-vector $\bar n$ which gives the direction of the final jet containing the observed photon. Since our cross section of interest is spherically symmetric (it is an $S$-wave annihilation), without loss of generality we take $n= (1,0,0,1)$.  We obtain the relevant SCET modes for our kinematics by projecting momenta along $n$, the conjugate null vector $\bar n = (1,0,0,-1)$, and the directions perpendicular to $n$ and $\bar n$.  The fields we keep are soft, with momentum scaling, $(\bar n \cdot p, \, n \cdot p, \, p_\perp) \sim  M_{\chi}(\lambda,\lambda,\lambda)$ and collinear with $p\sim M_{\chi}(1,\lambda^2,\lambda)$.  We work in units of $M_\chi$ which is the largest scale in the problem.  With this set of modes, the effective field theory is known as SCET-II.\footnote{In SCET-I, the ``soft'' modes have strictly smaller virtuality than the collinear ones, $p\sim (\lambda^2,\lambda^2,\lambda^2)$, and for this reason are sometimes called ultrasoft.} The collinear modes compose the jet in which the observed photon is produced. SCET is invariant under two distinct gauge symmetries, collinear (in however many collinear directions there are, which is just one in our case) and soft. This soft transformation is, in general, distinct from the soft or ultrasoft transformations in the NRQCD sector.
\begin{figure}
\centerline{\scalebox{.5}{\includegraphics{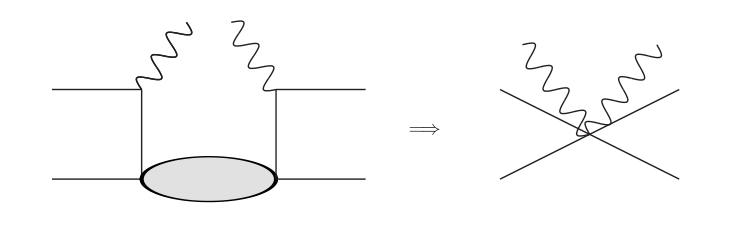}}}
\vskip-0.2cm
\caption[1]{Integrating out the final state - unobserved- jet. The open curly lines correspond to the observed photon which is accompanied 
by any amount of collinear radiation.}
\label{semi} 
\end{figure}

We build a factorization theorem at the level of the amplitude squared. This entails performing an operator product expansion
at scales $\gg \lambda M_\chi$, integrating out all final state particles except the final state jet containing {\it observed} photons and accompanying unobserved particles . 
This is illustrated in Figure \ref{semi}. Note that even if the balancing jet is composed of only one particle and has vanishing
invariant mass, the integration over the final state phase space shrinks the propagator to a point in the reduced Feynman
diagram (\ref{semi}). We thus match the full theory onto a set of operators with six fields, corresponding to the
incoming  winos and the outgoing, collinear, photon.  By taking matrix elements of these operators with external WIMP states, we obtain the matrix element-squared for semi-inclusive annihilation to $\gamma+X$, our observable of interest.
The minimal operator basis in the effective theory that we can write down is
\bea
O_1&=& \left( \bar  \chi \gamma^5 \chi \right) \vacproj \left( \bar \chi \gamma^5 \chi \right) B^{\mu A \perp} B_\mu^{ A \perp}\nn\\
O_2&=& \frac{1}{2}\Big\{\left( \bar \chi \gamma^5 \chi \right) \vacproj \left( \bar \chi_A \gamma^5 \chi_B \right) \nn \\
&&+ \left(\bar \chi_A \gamma^5 \chi_B \right) \vacproj \left( \bar \chi \gamma^5 \chi \right) \Big\} B_\mu^{\perp A} B^{\mu B \perp}\nn\\
O_3&=& \left( \bar \chi_C\gamma^5 \chi_D \right) \vacproj \left( \bar \chi_D \gamma^5 \chi_C \right) B^{\mu A \perp} B_\mu^{ A \perp}\nn\\
O_4&=& \left( \bar \chi_A \gamma^5 \chi_C \right) \vacproj \left( \bar \chi_C \gamma^5 \chi_B \right) B_\mu^{\perp A} B^{\mu B \perp},
\label{ops}
\eea
where we use the vacuum insertion approximation in the WIMP sector, which is valid up to $\mo(v^2)$ corrections.  Henceforth, we drop the explicit vacuum projector.  Implicitly, there is also a projection onto a single-photon state between the $B^{\mu \perp}$ fields i.e.
\beq
 B_\mu^{\perp A} B^{\mu B \perp}\equiv \sum_X  B_\mu^{\perp A}\mid \gamma+X\rangle\langle \gamma+X\mid B^{\mu B \perp}.
\eeq
where X contains the accompanying particles in the collinear jet.
All operators which arise in the  matching can be reduced to one of these four using
the Majorana condition.  The only relevant nonrelativistic bilinear is $\bar \chi \gamma_5 \chi$. The spin one operators are irrelevant 
since Fermi statistics would lead to an antisymmetric SU(2) initial state, and we are interested
in the annihilation of two neutral particles.  Furthermore, P-wave annihilation is velocity suppressed. We have also used the definition
\beq
B^{A \perp}_\mu \equiv f^{ABC}\, W_n^T (D^\perp_\mu)^{BC} \, W_n ,
\eeq
where the $\perp$ symbol implies the component perpendicular to the large light cone momentum $\bar n \cdot p$, where $\bar n^\mu=(1,0,0,-1)$ and $D^\perp_\mu$ is the covariant derivative
in the collinear sector (for details see \cite{SCET}).
This field interpolates for a collinear gauge boson and is invariant under collinear gauge transformations
due to the placement of the two collinear adjoint Wilson lines defined by
\beq
W_n^{BC}= P(e^{ g\int_{-\infty}^0 n \cdot A^A_n(n\lambda )f^{ABC} d\lambda}).
\eeq

For a general kinematic configuration, the $\chi$ fields do not transform under
SCET-II soft gauge transformations, since such a transformation would throw the
$\chi$ field off-shell. Thus, naively it appears that these
operators are trivially soft gauge invariant. Indeed this is true, but it does not
mean that the soft
mode does not play a role in this process.
The soft contribution to  the operators arises after one integrates out the
off-shell intermediate states which arise when softs couple to collinears.
The question then becomes, where should the soft Wilson lines be inserted into our
operators?
We could perform a matching calculation, however there is a much simpler method
which we call
the ``method of descent", developed in \cite{descent}. Here we will use a
variation of those arguments.

The idea is to choose a kinematic scenario in which the the invariant masses of the
external states are such that the soft momenta in  SCET-II  and the US modes in the NRQCD
sector have the same scaling. We also raise the virtuality of the collinear modes  so that
soft radiation leaves the collinear lines on-shell.\footnote{When there is a
hierarchy between the invariant masses of the soft and collinear momenta the theory
is usually called SCET-I and the soft fields are called ultrasoft. For the sake of
clarity we will call all non-collinear fields in SCET  ``soft''.}
In this scenario soft gauge invariance uniquely fixes the positions of the Wilson
lines.
The invariant mass of the external states is then lowered to its physical value,
keeping
the soft Wilson lines fixed. 

To apply this methodology to the case at hand we tune $M_\chi$ such that
$M_\chi v^2 \sim M_W$. This tells us that the ultrasoft modes in NRQCD scale as $M_{\chi}(\lambda,\lambda,\lambda)$. Simultaneously, we raise the virtuality of the collinear modes to be of
order $p_c^2\sim M_\chi M_W$. 
This would be appropriate if we were to, say, measure the jet mass and not the photon
energy, in which
case the color structure of the operator basis would remain as in Eq.~\ref{ops}. What this means is that we are writing down operators in a SCET-I theory with the expansion parameter $\lambda' = \sqrt{M_{W}/M_{\chi}}$ which exists at a higher scale $M_{\chi} \lambda'$. The ultrasoft modes in this EFT scale as $M_{\chi}(\lambda'^2,\lambda'^2,\lambda'^2)$ which is the same as the US modes of NRQCD in this limit.  Thus, US can communicate between the NR matter fields
and the collinear modes of SCET because such interactions leave both modes on shell. This apparent breakdown of factorization is
remedied by performing a BPS \cite{SCET} field redefinition
\bea
\chi &\rightarrow& S_v \chi \nn \\
B &\rightarrow& S_nB,
\eea
where there are two types of path ordered adjoint soft Wilson lines $S_v$ and $S_n$ defined by
\beq
S_{(v,n)bc}=P[ e^{g \int_{-\infty}^0 (v,n) \cdot A^a((v,n)\lambda)f^{abc} d\lambda}].
\label{eq:wilsonline}
\eeq
where the field $A^a$ supports modes which scale as $M_{\chi}(\lambda'^2,\lambda'^2,\lambda'^2)$.
This field redefinition decouples the US and collinear fields at the level of the action 
and dresses  the operators such 
that $O_2$ and $O_4$ become
\bea
O_2&=& \frac{1}{2}\Big\{(\bar \chi \gamma^5\chi) (\bar \chi_{A^\prime} \gamma^5 \chi_{B^\prime}) + (\bar \chi_{A^\prime} \gamma^5 \chi_{B^\prime}) (\bar \chi \gamma^5\chi) \Big\} B^{\tilde A} B^{\tilde B} \nn\\
&&S_{v A^\prime A}^\top \, S_{vB B^\prime} \, S_{n  \tilde A A}^\top \, S_{n B \tilde B}\nn\\
O_4&=& (\bar \chi_{A^\prime} \gamma^5 \chi_C)  (\bar \chi_C \gamma^5 \chi_{B^\prime})  B^{\tilde A} B^{\tilde B} \, S_{v A^\prime A}^\top \, S_{vB B^\prime} \, S_{n  \tilde A A}^\top \, S_{n B \tilde B}. 
\eea
The operators $O_1$ and $O_3$ receive no soft corrections.
We now continuously deform   $M_\chi$ back to its physical value and descend from the SCET-I theory at the scale $M_{\chi}\lambda'$ to the SCET-II theory at the scale $M_{\chi}\lambda$.  
In doing so, the soft fields retain their invariant mass of order $M_W$, 
and the soft Wilson lines remain fixed by continuity. 

The annihilation spectrum may be written as
\bea
\label{cross}
&& \frac{1}{E_\gamma}  \frac{d\sigma}{dE_\gamma} = \frac{1}{4M_\chi^2 v} \langle 0 | O^a_s | 0 \rangle \Bigg[ \int d n \cdot p \, \Bigg\{ C_2(M_\chi, n \cdot p) \langle p_1 p_2 \mid \frac{1}{2}\Big\{(\bar \chi \gamma^5\chi) \, (\bar \chi_{A^\prime} \gamma^5 \chi_{B^\prime}) \nn \\
&+& (\bar \chi_{A^\prime} \gamma^5 \chi_{B^\prime}) (\bar \chi \gamma^5\chi) \Big\}(0) \mid p_1 p_2 \rangle + C_4(M_\chi, n \cdot p) \langle p_1 p_2 \mid (\bar \chi_{A^\prime} \gamma^5 \chi_C) \, (\bar \chi_C \gamma^5 \chi_{B^\prime}) (0) \mid p_1 p_2 \rangle \Bigg\} F^\gamma_{\tilde A \tilde B}\left( \frac{2E_\gamma}{n \cdot p} \right) \Bigg] \nn\\
&&+  \Bigg[ \int d n \cdot p \,\Bigg\{ C_1(M_\chi, n \cdot p)\langle p_1 p_2 \mid (\bar \chi \gamma^5 \chi) \, (\bar \chi \gamma^5 \chi) (0) \mid p_1 p_2 \rangle +C_3(M_\chi, n \cdot p) \nn \\
&\times& \langle p_1 p_2 \mid (\bar \chi_C \gamma^5 \chi_D) \,  (\bar\chi_D \gamma^5 \chi_C) (0) \mid p_1 p_2 \rangle\Bigg\} F_\gamma\left( \frac{2E_\gamma}{n \cdot p} \right)  \Bigg],
\eea
where  
\bea
O^a_s= S_{v A^\prime A}^T S_{vB B^\prime}S_{n  \tilde A A}^TS_{n B \tilde B}
\eea
and $F^\gamma_{\tilde A \tilde B}$ is a fragmentation function defined by
\bea
F^\gamma_{\tilde A \tilde B}\left( \frac{n\cdot k}{n\cdot p} \right) &=& \int \frac{ dx_-}{2\pi}e^{in \cdot p x_-}\langle 0 \mid B^{\perp \mu}_{\tilde A}(x_-) \mid \gamma(k_n)+X_n \rangle \nn \\
&\times& \langle \gamma(k_n)+X_n \mid  B^\perp _{\mu \tilde B}(0)  \mid 0 \rangle ,
\label{eq:fragdef}
\eea 
and $F_\gamma = F^\gamma_{\tilde A \tilde B} \delta_{\tilde A \tilde B}$.
Note that this is an unusual fragmentation function in that we are measuring states which are not gauge singlets.  However, electroweak symmetry breaking makes 
this physically meaningful since it is trivial for observers to agree on the appropriate, EM-preserving gauge.

$C_{1\textendash 4}$ are the matching coefficients that give the probability for the dark matter to annihilate and create a photon with momentum $ n \cdot p$.
$F^\gamma$ is the canonical fragmentation function giving the probability of an initial photon with momentum $k$ to yield
a photon with momentum fraction $n\cdot k /n \cdot p$ after splitting. Since the contribution in Eq.~(\ref{cross}) proportional to $F^\gamma$
is not sensitive to the nonsinglet nature of the initial state, it will only contribute large double logs from mixing with $O_{2,4}$. 

In writing down Eq.~\ref{eq:wilsonline}, we factorized the collinear and soft fields, as the total Hilbert 
space of the system is a tensor product of the soft and collinear sector. In general, none of the NRQCD modes can interact with the SCET mode without throwing them off-shell, thus leading to power suppressed interactions. Of course, the size of the power corrections will be dictated by this offshellness, but independently of the system's details, these interactions will not lead to large double logs.

\section{The Role of the Higgs}
\label{sec:higgs}

Given that the Higgs mass is of order the weak scale, in principle we should consider
both soft and collinear Higgs emissions in our factorization theorem. However, as we shall
now show, the couplings of both collinear and soft Higgses are power suppressed. Assuming that the
Higgsino is sufficiently heavier than the winos we may neglect the couplings of the Higgs to
the nonrelativistic sector of the Lagrangian. In cases with light Higgsinos, the potential is 
affected, as discussed in \cite{Beneke:2014gja}.

The coupling of the Higgs to the gauge bosons is given by
\beq
S_H \sim \int d^4x \, W_\mu W^\mu H^\dagger H.
\eeq
Let us consider the coupling of a soft Higgs to collinear gauge bosons. 
Recall that for a collinear gauge field with large momentum $n \cdot p$, the
polarizations scale as
\beq
(n\cdot W \sim 1, \bar n \cdot W \sim \lambda^2, W_\perp \sim \lambda),
\eeq
where the power counting parameter is $\lambda \sim M_W/M_\chi$.
The soft higgs field scales as $\lambda$, while the measure scales as
$\lambda^{-3}.~$\footnote{The scaling of the measure is based on the support of the
fields in momentum space, or equivalently by power counting the delta (see for instance \cite{TASI}). }
Thus the coupling of a soft Higgs to a collinear jet is down by one factor of $\lambda$.

Now suppose we are interested in the coupling of a collinear Higgs to a collinear gauge boson.
In principle this could, as in the case of soft Higgs emission, lead to non-analyticity that
must be reproduced by the effective theory.  The coupling of collinear Higgs with collinear gauge
bosons (in the same light cone direction) is leading order since the measure will now scale as $\lambda^{-4}$.
Such interactions will be written in the effective theory as 
\beq
S^n_{HW}= \int d^4x g^2(B^A_{n\mu} \tau^A H_n)^T (B^{B\mu}_n \tau^B H_n) \sim O(1).
\eeq
This interaction will generate running in the fragmentation functions at subleading orders only.

It is interesting to ask whether or not the emission of a collinear Higgs in the $n^\prime$ direction
off of a particle in the $n$ direction can generate leading order interactions. The analogous emission of
an arbitrary number of gauge bosons in the matching procedure is what builds up the Wilson lines in the effective theory.
Thus one might suspect that the analogous mechanism should  {\bf not} occur in the case of Higgs emission given
that Wilson lines are already there to insure gauge invariance.  This suspicion is proven true by noting a crucial
distinction between collinear Higgs and  gauge boson emission.  The Higgs field scales as $\lambda$ whereas the
gauge boson component $n\cdot A\sim 1$. This is why one can emit an arbitrary number of gauge bosons 
in an amplitude without power suppression. Thus when matching, the emission of a Higgs in the $n^\prime$
direction off a parton in the $n$ direction is power suppressed and will be neglected.

\section{Calculating the Anomalous Dimension}
\label{sec:ad}

\begin{figure}[ht!]
\centerline{\scalebox{0.5}{\includegraphics{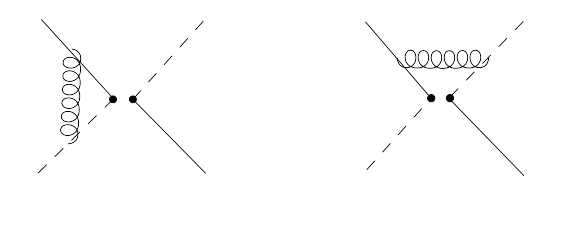}}}
\caption[1]{Diagrams which contribute rapidity divergences to the soft factor. The dashed/solid line represents
the time/light-like Wilson line. }
\label{soft} 
\end{figure}

Much of the analysis in this section we first presented in \cite{Baumgart:2014vma}.  To calculate the anomalous dimensions we first introduce an operator basis in the collinear and soft sectors
\bea
O_s^a&=&S_{v A^\prime A}^T S_{vB B^\prime}S_{n  \tilde A A}^TS_{n B \tilde B}~~~~~~~~~O_s^b=\mathbb{1} \, \delta_{\tilde A \tilde B} \delta_{A^\prime B^\prime}
\nn \\
O_c^a &=& B^\perp_{\tilde A}\mid \gamma(k_n)+X_n \rangle \langle \gamma(k_n)+X_n \mid B^\perp_{\tilde B}\nn\\
O_c^b&=& B^\perp_D \mid \gamma(k_n)+X_n \rangle  \langle \gamma(k_n)+X_n \mid B^\perp_D \delta_{\tilde A \tilde B}.
\eea
where there is an implicit sum over the polarizations of the photon.
The operator $O_s^b$ has a trivial structure and hence does not receive radiative corrections, meaning its anomalous dimension is 0. Since it is a color singlet operator, the real and virtual poles cancel in the corrections to $O_c^b$, thus its anomalous dimension is 0 as well.  At tree level, the vacuum matrix element of the operator $O_s^a$ is simply $\delta_{ A^\prime \tilde A} \delta_{B^\prime \tilde B}$.
At one loop, the diagrams that contribute to this matrix element are shown in Fig.~\ref{soft}.
The soft and collinear modes have the same virtuality and hence the divergences that arise from the factorization of the soft sector from the collinear cannot be regulated by dimensional regularization, which respects boost symmetry. Hence, we need to introduce a rapidity regulator, which manifestly breaks boosts \cite{RRG}. This requires a corresponding factorization scale which we call $\nu$. Using this formalism for the soft sector gives us
\bea
\langle 0|O_s^a|0 \rangle  =\!  \delta_{ A^\prime \tilde A} \delta_{B^\prime \tilde B} \!+ \{\delta_{\tilde A \tilde B} \delta_{A^\prime B^\prime} \!-\! 3\, \delta_{ A^\prime \tilde A} \delta_{B^\prime \tilde B}\}\frac{g^2}{4\pi^2}\left[2 \log(\frac{\nu}{\mu}) \log(\frac{\mu}{M_W})\!+\log^2(\frac{\mu}{M_W})\right].
\eea
Thus, we see that even though $O_s^a$ itself has a nontrivial color structure, at one loop it generates a color singlet piece in addition to nonsinglet.

\begin{figure}[ht!]
\centerline{\scalebox{0.6}{\includegraphics{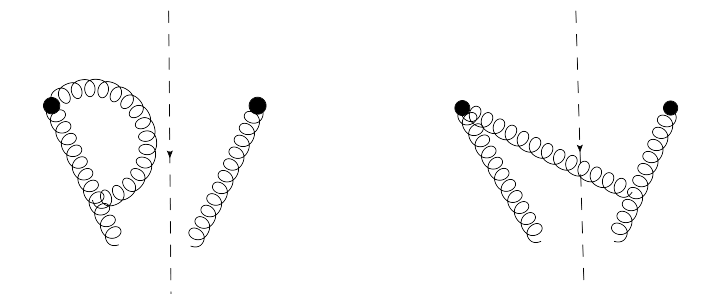}}}
\vskip-0.2cm
\caption[1]{The two diagrams which lead to rapidity divergences in the fragmentation function.
The dashed line represents the cut throughout which final states pass. The solid dot represents
the gauge invariant field strength $B_\mu^\perp$.}
\label{coll} 
\vskip-0.3cm
\end{figure}

We can do a similar calculation for the vacuum matrix elements in the collinear sector ({\it cf.}~Fig.~\ref{coll}).
\bea
\langle 0|O_c^b|0 \rangle &=& 2\delta_{\tilde A \tilde B}\nn\\
\langle 0|O_c^a|0 \rangle  &=&  2\delta_{ \tilde A 3} \delta_{\tilde B 3} + 2\{\delta_{\tilde A \tilde B}-3\delta_{ \tilde A 3} \delta_{\tilde B 3}\}\frac{g^2}{4\pi^2}\{2 \log(\frac{M_{\chi}}{\nu}) \log(\frac{\mu}{M_W})\}
\eea
These operators clearly mix within their respective sectors and we can define anomalous dimension matrices for the scales $\mu$ and $\nu$.

\beq \mu \frac{d}{d\mu}\left(\begin{array}{c} O^{c,s}_a \\ O^{c,s}_b \end{array}\right)=\left( \begin{array}{cc} \gamma^{c,s}_{\mu,aa} & \gamma^{c,s}_{\mu,ab}  \\ 0 & 0 \end{array} \right) \left(\begin{array}{c} O^{c,s}_a \\ O^{c,s}_b \end{array}\right). \eeq
\beq \nu \frac{d}{d\nu}\left(\begin{array}{c} O^{c,s}_a \\ O^{c,s}_b \end{array}\right)=\left( \begin{array}{cc} \gamma^{c,s}_{\nu,aa} & \gamma^{c,s}_{\nu,ab}  \\ 0 & 0 \end{array} \right) \left(\begin{array}{c} O^{c,s}_a \\ O^{c,s}_b \end{array}\right). \eeq
The anomalous dimensions are thus given by
\bea
\gamma^c_{\mu,aa} &=& \frac{3g^2}{4\pi^2} \log(\frac{\nu^2}{4M^2_\chi}), \;\;\; \gamma^s_{\mu,aa} = \frac{-3g^2}{4\pi^2} \log(\frac{\nu^2}{\mu^2}), \nn \\
\gamma^c_{\mu,ba} &=& \frac{-g^2}{4\pi^2} \log(\frac{\nu^2}{4M^2_\chi}), \;\; \gamma^s_{\mu,ba} = \frac{g^2}{4\pi^2} \log(\frac{\nu^2}{\mu^2}). 
\label{ads}
\eea
\bea
\gamma^c_{\nu,aa} &=& \frac{3g^2}{4\pi^2} \log(\frac{\mu^2}{M_W^2}), \;\;\; \gamma^s_{\nu,aa} = \frac{-3g^2}{4\pi^2} \log(\frac{\mu^2}{M_W^2}), \nn \\
\gamma^c_{\nu,ba} &=& \frac{-g^2}{4\pi^2} \log(\frac{\mu^2}{M_W^2}), \;\; \gamma^s_{\nu,ba} = \frac{g^2}{4\pi^2} \log(\frac{\mu^2}{M_W^2}). 
\label{ads2}
\eea
Since the rapidity regulator was needed to handle the divergence that arose from our artificial cut between the soft and collinear sector, any trace of it, including dependence on the scale $\nu$ should vanish when we combine soft and collinear results.  Therefore, 
the fact that the $\nu$ anomalous dimension matrices in the soft and collinear sector are equal and opposite and that $\nu$-dependence cancels when we sum soft and collinear $\mu$ anomalous dimensions provides a powerful cross check of our effective theory calculation.

The soft and collinear  sectors have no large logs if we choose the $(\mu,\nu)$ scales to be $(M_W,M_W)$ and
 $(M_W,M_\chi)$ respectively. At leading double log accuracy we can resum all of the relevant terms
 by choosing $\mu=M_W$. In this case all the large logs reside in the renormalized parameter $C_{i}(\mu=M_W)$
 and the rapidity running may be neglected.  We can read off the running of the hard matching coefficients $C_{1 \textendash 4}$ of the operators in Eq.~\ref{ops} by imposing that the cross section be RG invariant

\bea
\mu \frac{d}{d\mu}C_{2,4}(\mu) &=& - (\gamma^c_{\mu,aa}+\gamma^s_{\mu,aa}) C_{2,4} \nn \\
\mu \frac{d}{d\mu}C_{1,3}(\mu) &=& - (\gamma^c_{\mu,ba}+\gamma^s_{\mu,ba}) C_{2,4}. 
\label{wilsonrg}
\eea
Notice that the RHS of Eq.~\ref{wilsonrg} is independent of the rapidity scale as it must be.

To present a model independent form we have used the fact  that the tree level result for $\chi^0 \chi^0 \rightarrow \gamma + \gamma/Z$ must vanish  in which case there is only one independent matching coefficient. This gives us that $C_1=C_4$, $C_3=0$ and $C_2=-2C_1$ for matching at the high scale $M_{\chi}$. Using these boundary conditions we can solve for the Wilson coefficents at the low scale $\mu$ $\sim$ $M_{W}$. 
\bea
 C_2(M_W)&=& \exp \left[ -\frac{3g^2}{4\pi^2}\log^2(M_\chi/M_W)\right] C_2(M_\chi) = -2 \exp\left[ -\frac{3g^2}{4\pi^2}\log^2(M_\chi/M_W)\right] C_1(M_\chi) \nn \\
 C_1(M_W) &=&C_1({M_\chi})+\frac{1}{3}\left\{1-\exp\left[-\frac{3g^2}{4\pi^2}\log^2(M_W/M_\chi)\right]\right\}C_2(M_\chi) \nn\\
&=& \left\{\frac{1}{3} +\frac{2}{3}\exp\left[-\frac{3g^2}{4\pi^2}\log^2(M_W/M_\chi) \right] \right\} C_1({M_\chi})
\label{eq:wilsonrunone}
 \eea
Similarly for $C_3$ and $C_4$
  \bea
 C_4(M_W)&=& \exp\left[ -\frac{3g^2}{4\pi^2}\log^2(M_\chi/M_W)\right] C_1(M_\chi) \nn \\
 C_3(M_W) &=&\frac{1}{3}\left\{ 1-\exp\left[-\frac{3g^2}{4\pi^2}\log^2(M_W/M_\chi)\right] \right \} C_1(M_\chi)
\label{eq:wilsonruntwo}
 \eea

\begin{figure}[ht]
\centerline{\scalebox{0.6}{\includegraphics{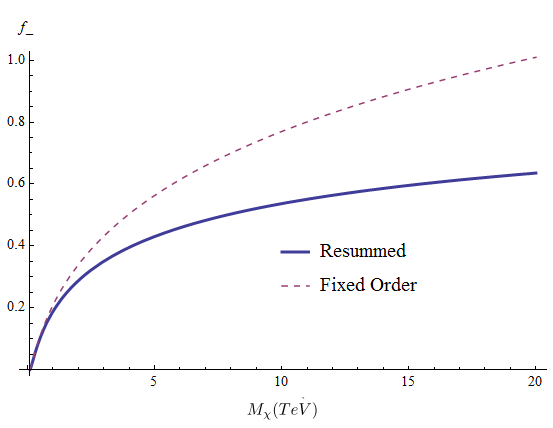}}}
\vskip-0.25cm
\caption{The resummed Sudakov factor $f_-$ as a function of the neutralino mass $M_{\chi}$. }
\label{fm} 
\vskip-0.5cm
\end{figure}

\begin{figure}[ht]
\centerline{\scalebox{0.6}{\includegraphics{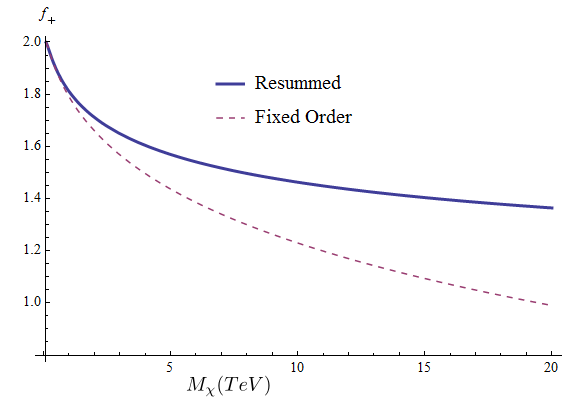}}}
\vskip-0.3cm
\caption{The resummed Sudakov factor $f_+$ as a function of the neutralino mass $M_{\chi}$. }
\label{fp} 
\vskip-0.5cm
\end{figure}

By running the Wilson coefficients, we resum all the leading double log contributions. The cross section can now be obtained by evaluating the effective theory matrix elements at their natural scale $\mu \sim M_W$. Here, all matrix elements are merely their values at tree level.
\bea
F^\gamma_{\tilde A \tilde B}(n\cdot k/n\cdot p) &\equiv& \int \frac{ dx_-}{2\pi}e^{i n \cdot p \, x_-} 
\langle 0 | O_c^a | 0 \rangle \nn \\
&=& \int \frac{ dx_-}{2\pi}e^{i(n \cdot p x_--n \cdot k_n x_-)} \delta_{\tilde A 3} \delta_{\tilde B 3}\eta^{\mu \nu} \sum_{pols} \epsilon_{\mu \perp}(k_n)\epsilon^{*}_{\nu \perp}(k_n)\nn\\
&=& 2 \delta_{\tilde A 3} \delta_{\tilde B 3} \delta(n \cdot p -n \cdot k_n) \nn \\
\langle 0 \mid O_s^a \mid 0 \rangle &=& \delta_{A^\prime \tilde A}  \delta_{B^\prime \tilde B}
\eea
At the low scale, we are working in the broken theory, where the mass eigenstates are the neutralino $\chi^0$ and the charginos, $\chi^{\pm}$, which are defined as 
\bea
\chi^0= \chi^3\nn\\
\chi^{\pm}= \frac{1}{\sqrt{2}}(\chi^1 \mp i\chi^2)
\eea
Anticipating the form of the Wilson coefficient at the scale $M_{\chi}$, we pull out a delta function $\delta(E_\gamma-M_\chi)$. Thus, utilizing the simplifications above along with the new basis we obtain the final form of the cross section up to corrections in the relative velocity \cite{Baumgart:2014vma}.

\bea
\label{final}
\frac{1}{E_\gamma}\frac{d\sigma}{dE_\gamma} &=& \frac{C_1(\mu=E_{\gamma})}{4M^2_{\chi} \, v}\delta(E_\gamma-M_\chi) \left[ \frac{4}{3}f_- \mid \! \psi_{00}(0)\!\mid^2+4 f_+\mid \! \psi_{+-}(0)\!\mid^2\nn \right. \\
&+& \left. \frac{4}{3}f_-(\psi_{00}\psi_{+-}+{\rm h.c.}) \right]
\label{eq:diffrate}
\eea
where  $f_\pm \equiv 1\pm\exp[- \frac{3\alpha_W}{\pi}\log^2(\frac{M_W}{E_{\gamma}})]$, which we plot in Figs.~\ref{fm} and \ref{fp}.  Because $f_+ > f_-$ by a factor of a few throughout our range of interest and the numerical prefactor and Sommerfeld factors, as we will discuss in the Sections \ref{sec:se}, maintain this modest hierarchy, one can obtain a quick estimate of the effects of radiative corrections and resummation to the rate by looking at Fig.~\ref{fp}.
We define the wavefunctions as 
\begin{eqnarray}
\psi_{00}(0) &=& \langle 0 | (\chi^0)^\top  i \sigma^2 \chi^0 |\chi^0 \chi^0 \rangle_S \nn \\
\psi_{\pm}(0) &=& \langle 0 |  (\chi^-)^\top i \sigma^2 \chi^+ | \chi^0 \chi^0 \rangle_S ,
\label{eq:wfxn}
\end{eqnarray}
where $|\chi^0 \chi^0 \rangle_S = \frac{1}{\sqrt{2}} (|\chi^0_{\uparrow}(p_1) \chi^0_{\downarrow}(p_2)\rangle-|\chi^0_{\downarrow}(p_1) \chi^0_{\uparrow}(p_2)\rangle)$, and here after we drop the ``$S$'' subscript since our nearly-static, annihilating Majorana fermions are automatically in the spin singlet.

Writing in terms of two component fields makes the singlet structure manifest.
This gives us 
\begin{eqnarray}
\bar{\chi}^0 \gamma^5 \chi^0 &=& 2 (\chi^0)^T i\sigma^2 \chi^0 \nn\\
\bar{\chi}^+ \gamma^5 \chi^+ &=& 2(\chi^-)^T i\sigma^2 \chi^+ ,
\end{eqnarray}
where the two-component fermions on the RHS are the large-component subset of the four-component spinors on the LHS in the nonrelativistic limit.  In order to fix the Wilson coefficient $C_1$, we match onto the tree level annihilation cross section of a spin singlet chargino state  $ \frac{1}{\sqrt{2}} (|\chi^+_{\uparrow}(p_1) \chi^-_{\downarrow}(p_2)\rangle-|\chi^+_{\downarrow}(p_1) \chi^-_{\uparrow}(p_2)\rangle)$.
The leading order cross section to $\gamma + X$ $(\chi^+ \chi^- \rightarrow \gamma \gamma + \frac{1}{2} (\chi^+ \chi^- \rightarrow \gamma Z))$ is given as 
\begin{eqnarray}
\sigma v = \frac{\pi \alpha_W^2 \sin^2\theta_W}{M_{\chi}^2}
\end{eqnarray}
From the effective theory description, we calculate this particular cross section in the notation of Eq.~\ref{eq:diffrate} as 
\bea
\frac{1}{E_\gamma}\frac{d\sigma}{dE_\gamma} &=& \frac{C_1(\mu=E_{\gamma})}{4M^2_{\chi} \, v}\delta(E_\gamma-M_\chi) (4 f_+ | \langle0| (\chi^-)^T i\sigma^2 \chi^+ |\chi^+ \chi^- \rangle_S|^2)
\label{eq:match}
\eea
At tree level, $f_+ =2$  and $\langle 0| (\chi^-)^T i\sigma^2 \chi^+ |(\chi^+ \chi^-)_S \rangle= \sqrt{2} M_{\chi}$.\\
which gives us 
\bea
\sigma v = C_1(\mu=M_{\chi}) 2M_{\chi}
\eea
Comparing the two results we fix 
\beq
C_1(M_{\chi}) = \frac{\pi \alpha_W^2 \sin^2\theta_W}{2M_{\chi}^3}.
\eeq

\section{Sommerfeld Enhancement}
\label{sec:se}

In order to quantify the semi-inclusive rate calculation, we need to determine the wavefunction-at-the-origin factors that enter our final, LL-resummed differential cross section in Eq.~\ref{eq:diffrate}.  The wavefunctions themselves are defined in Eq.~\ref{eq:wfxn} and can be computed in principle in the nonrelativistic effective theory by summing the ladder exchange of electroweak gauge bosons between winos to all orders.  Fortunately, this is equivalent to the operationally simpler task of solving the Schr\"odinger equation for our two, two-body states $| \chi^0 \chi^0 \rangle$ and $| \chi^+ \chi^- \rangle$ in the presence of the electroweak potential \cite{Hisano:2004ds,Iengo:2009ni,Cassel:2009wt}.  In Appendix \ref{app:pot}, we detail the process of obtaining this potential from the underlying quantum field theory.  Since it contains Coulomb, Yukawa, and mass-shift pieces and is off-diagonal for the two states, we solve it numerically.  As expected for slowly moving particles in the presence of an attractive potential, we find Sommerfeld enhancement for the annihilation, that for some regions of $M_{\chi}$ is orders of magnitude above the perturbative rate.  

Taking into account appropriate state normalization, the Schr\"odinger potential is
\bea
V(r) = \left( \begin{array}{cc}
2\delta M -\frac{\alpha}{r} - \alpha_W c_W^2 \frac{ e^{-m_Z r}}{r} & -\sqrt{2}\alpha_W \frac{e^{-m_W r}}{r}   \\
-\sqrt{2}\alpha_W  \frac{e^{-m_W r}}{r}  &  0 
\end{array} \right),
\label{eq:potl}
\eea
where $\delta M \equiv M_{\chi^+} - M_{\chi^0}$.  For numerical analysis, we use $\delta M$ = 0.17 GeV, which is its value over much of MSSM parameter space.  To solve the system, we also need appropriate boundary conditions for our wavefunctions $\psi_{1} \equiv \langle r| \chi^0 \chi^0 \rangle $ ($\psi_{2} \equiv \langle r| \chi^+ \chi^- \rangle $), where $r$ is the relative distance between the two particles in the state, and we always work in the center of mass frame. The total wavefunction for our two state system is $\boldsymbol{\psi}^\top = (\psi_1 \; \psi_2)$.  We are ultimately interested in the annihilation of the neutral state, which is controlled by physics at length scales $\sim \frac{1}{M_\chi} \ll \frac{1}{M_W}$, and is therefore quantified by the wavefunction-at-the-origin, $\psi_{1,2}(0)$.  We thus see that for the boundary condition of an incoming neutral, spin-singlet state, in the notation of Eqs.~\ref{eq:diffrate} and \ref{eq:wfxn}, 
\bea
\psi_{1}(0) &=& \psi_{00}(0)/(2\sqrt{2}M_\chi) \nn \\ 
\psi_{2}(0) &=& \psi_{\pm}(0)/(2M_\chi), 
\label{eq:wfrat}
\eea
where the numerical factors account for the difference between our definition in Eq.~\ref{eq:wfxn} and the interpolating fields for the two-body states and the tree-level normalization of the matrix element below Eq.~\ref{eq:match}.  The latter can be found by comparing the potential we use for Schr\"odinger evolution (Eq.~\ref{eq:potl}) with the nonrelativistic field theory interactions we obtain in Eq.~\ref{eq:qftpotl}.  Since the annihilation process is perturbative, we can calculate $\psi_{1,2}(0)$ by turning off the annihilation and solving the scattering problem for the electroweak potential.\footnote{In the literature the Sommerfeld enhancement is sometimes quantified as $S \equiv |\psi(0)|^2/|\psi^{(0)}(0)|^2$, the ratio of the potential-modified wavefunction at the origin to the wavefunction of a plane wave, $\psi^{(0)} = e^{ikz}$.  However, since $|\psi^{(0)}(0)|^2$ = 1, and our expression for the differential rate, Eq.~\ref{eq:diffrate}, is written explicitly in terms proportional to $\psi_{1,2}(0)$, our notation only refers to the wavefunctions.}  This is equivalent to our factorized effective field theory setup ({\it cf.}~Eq.~\ref{eq:diffrate}), where the wavefunctions are computed in the nonrelativistic effective theory, but the annihilation process is given by the perturbative, high-energy Wilson coefficient. 

To proceed, we adopt the general analysis of \cite{Slatyer:2009vg,Beneke:2014gja}, and stick to the former's notation as much as possible, for our state of interest. As is generic for a scattering problem in the presence of a central potential, for our wavefunction,
\beq
\psi_{n}(r \rightarrow \infty) = c_n e^{i k_n z} + f_n(\theta) \frac{e^{i k_n r}}{r},
\label{eq:scattinf}
\eeq
where we recognize the incoming plane wave and scattered, outgoing radial wave.  The index $n$ labels the charged or neutral component of the two-body state, with $n=1$ being neutral.  Since our asymptotic, physical state is a pair of neutral winos with CM frame $E = \frac{M_\chi v^2}{4}$, we demand that the incoming plane wave only be in the neutral component of the state, {\it i.e.}~$c_n = \delta_{n1}$, and 
\beq
k_n = \frac{M_\chi}{2}\sqrt{v^2 - 8 \frac{\delta M}{M_\chi}\delta_{n2}}.
\eeq
The presence of the mass shift, $\delta M$, causes the charged component of the state to decay exponentially at large $r$.  A generic wavefunction in a spherical potential can be decomposed in terms of a radial wavefunction $R_{kl}(r)$ and Legendre polynomials as,
\beq
\psi_{n}(r) = \sum_\ell A_{n\ell} \, P_\ell (\cos \theta) R_{n,k\ell}(r),
\label{eq:decomp}
\eeq
We wish to determine the unknown coefficient, $A_{n\ell}$, by matching to Eq.~\ref{eq:scattinf} as $r \rightarrow \infty$.  
For this, we expand the incoming plane wave into partial waves and also use the general form of the radial wavefunction for central potential scattering at long distances, 
\beq
R_{nkl}(r \rightarrow \infty) = b_{n\ell} \, \sin (k_n r -\ell \pi/2 +\delta_{n\ell})/r, 
\label{eq:rgen}
\eeq
where $b_{n\ell}$ is a constant and $\delta_{n \ell}$ is the ``phase shift.''  Strictly speaking, only the radial component of $\psi_{1}$ asymptotes to this form.  As mentioned above, the radial component of $\psi_{2}$ decays exponentially.  We can analytically continue $k_2$ and $\delta_{2\ell}$ though, to complex values to keep the decaying solution in the form of Eq.~\ref{eq:rgen}.  Formally,  $\delta_{2\ell}$ diverges to cancel the exponentially growing mode for complex $k_2$, but as we will see, we never need to input its value to determine Sommerfeld enhancement.  

To proceed with matching, it is useful to include the other, regular, linearly-independent solution to the Schr\"odinger equation (generically an $N$-component state has $N$ regular and $N$ irregular solutions, that latter will be useful for our numerical analysis as described below).  Since we were already considering the case of an incoming neutral state in Eq.~\ref{eq:scattinf}, $c_n = \delta_{n1}$, we can take the orthogonal solution to be that of an incoming charged state, $c_n = \delta_{n2}$.  Thus, anything with a $n$ index becomes a tensor with indices, $n,j$ (except $k_n$, since it is kinematic and thus invariant across solutions), where $j$ determines whether the incoming plane wave is neutral ($j=1$) or charged.  For $c$, we therefore get $c_{nj} = \delta_{nj}$, and $j$ iterates between an incoming $\chi^0 \chi^0$ state, what we were already considering, or an incoming $\chi^+ \chi^-$.  We can now define matrices, $C_{nj} \equiv c_{nj}/k_n$ and $(M_\ell)_{nj} \equiv e^{-i(\delta_\ell)_{nj}} (b_\ell)_{nj}$.  This allows us to find the matching coefficient in Eq.~\ref{eq:decomp},
\beq
A_\ell = i^\ell (2\ell+1)M_\ell^{-1}C,  
\label{eq:acoeff}
\eeq
where the first column of $A_\ell$ gives the coefficients for our state of interest.  

In the opposite limit, $R_{kl}(r \rightarrow 0) \sim r^\ell$, and since our only concern is the wavefunction at the origin, we need only keep track of the $S$-wave component.  As a further simplification, we define $\chi_k \equiv r R_{k,\ell=0}$, such that the Schr\"odinger equation reduces to 
\beq
-\frac{1}{M_\chi}\frac{d^2}{dr^2}\chi(r) + V(r) \, \chi(r) = \frac{k^2}{M_\chi}\chi(r).
\label{eq:redschro}
\eeq
Combined with our matching coefficient, this gives a matrix of Sommerfeld enhancement factors,
\beq
\psi_{nj}(0) = (\chi^\prime(0) M_0^{-1} C)_{nj},
\label{eq:sommphys}
\eeq
where $j$ labels the external state and $n$ determines whether the 00 or $\pm$ component undergoes perturbative annihilation.  

As setup currently though, we are on the hook for calculating the phase shifts, $\delta_{nj}$, and amplitudes, $b_{nj}$, of the asymptotic solutions.  To remove this difficulty and any subtleties about the decaying nature of the charged-component term, it is useful build the Wronskian with the other two linearly-independent solutions of Eq.~\ref{eq:redschro}, those that give irregular radial wavefunctions.  Denoting them as $(\tilde{\chi}_\ell)_{nj} (r)$, we demand
\bea
(\tilde{\chi}_\ell)_{nj} (r \rightarrow \infty) &=& T_{nj} e^{i k_n r} \nn \\
(\tilde{\chi}_\ell)_{nj} (r \rightarrow 0) &=& r^{-\ell} \delta_{nj}.
\eea
Dropping $\ell$, since we are only interested in the $S$-wave case, $W = \tilde{\chi}^\top \chi^\prime - \tilde{\chi}^{\prime\top} \chi$.  $W$ is invariant with respect to $r$, and so we evaluate it at 0 and $\infty$,
\bea
W(0) &=& \chi^\prime(0) \nn \\
W(r \rightarrow \infty) &=& \sum_{m} T_{mn} (M)_{mj} k_m \nn \\
&=& (T^\top C^{-1} M)_{nj},
\eea
where we have used that the $C$ matrix is diagonal, with elements $k_m^{-1}$.  Finally, we can get a simpler expression for Sommerfeld enhancement, since equating $W(0)=W(\infty)$ gives
\beq
(\chi^\prime(0) M^{-1} C)_{nj} = T^\top_{nj}. 
\label{eq:sommeq}
\eeq
However, the LHS is exactly what we obtained in Eq.~\ref{eq:sommphys}.  Thus, for numerical analysis, we only need to find the neutral component of the wavefunction at infinity, $\tilde{\chi}_{11}(r \rightarrow \infty)$ and $\tilde{\chi}_{12}(r \rightarrow \infty)$, where the different solutions correspond to imposing the boundary conditions 
\bea
\begin{array}{cc}
\tilde{\chi}_{11}(0) = 1 & \;\;\;\; \tilde{\chi}_{12}(0) = 0   \\
\tilde{\chi}_{21}(0) = 0  & \;\;\;\; \tilde{\chi}_{22}(0) = 1  
\end{array}  
\label{eq:bc}
\eea
Furthermore, since $\tilde{\chi}_{1i}(r \rightarrow \infty) \propto e^{ikr}$ we can extract the phase part, defining $\xi_{i} \equiv \tilde{\chi}_{1i} e^{-ikr}$ such that $\xi^\prime(\infty) = 0$, giving an even simpler numerical implementation.

\begin{figure}[ht!]
\centerline{\scalebox{0.5}{\includegraphics{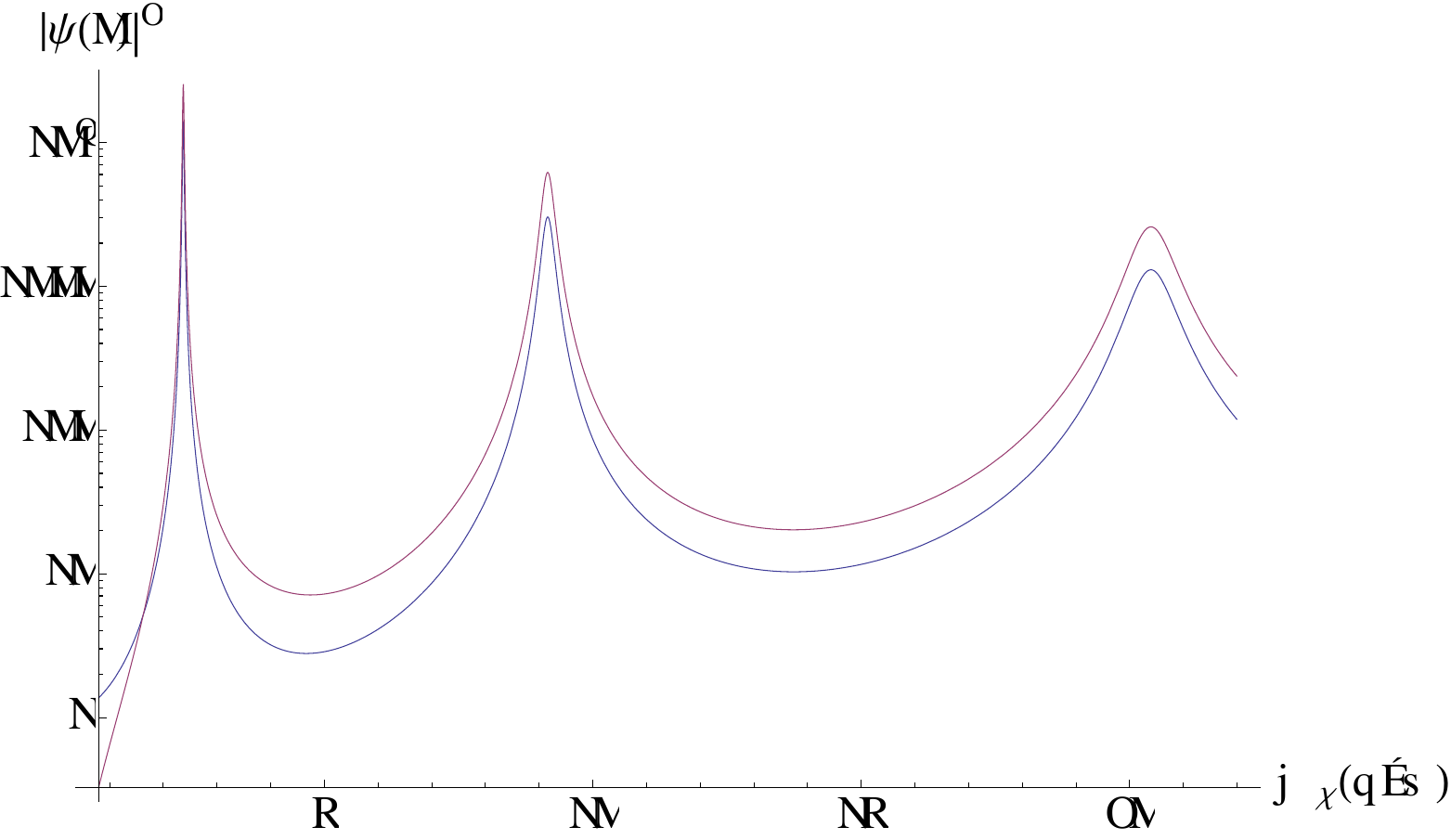}}}
\caption[1]{Sommerfeld enhancement factors, $|\psi_2(0)|^2$ (red) and $|\psi_1(0)|^2$ (blue) vs.~WIMP mass.  The former is promotional to $|\psi_{\pm}(0)|^2$ and the latter to $|\psi_{00}(0)|^2$, as shown in Eq.~\ref{eq:wfrat}.} 
\label{somm}
\end{figure}

We take the reduced Schr\"odinger equation, Eq.~\ref{eq:redschro}, and solve for $\xi_{i} (\equiv \tilde{\chi}_{1i} e^{-ikr})$ and $\tilde{\chi}_{2i}$ with the potential in Eq.~\ref{eq:potl} and boundary conditions at the origin given in \ref{eq:bc}.  Defining $x\equiv k r$, where $k=M_\chi v/2$, this gives the following coupled equations,
\bea
\xi^{\prime\prime}_i(x) + 2i \xi^\prime_i(x) &=& -\sqrt{2} \left( \frac{\alpha_W M_\chi}{x \, k}\right) e^{-(i+M_W/k)x} \tilde{\chi}_{2i}(x) \nn \\
\tilde{\chi}^{\prime\prime}_{2i}(x) &=& -\sqrt{2} \left( \frac{\alpha_W M_\chi}{x \, k} \right) e^{-(i+M_W/k)x} \xi_i(x) \nn \\
&&+ \left( \frac{2 M_\chi \delta M}{k^2} - \frac{\alpha M_\chi}{x \, k} -  \left( \frac{\alpha_W M_\chi c_W^2}{x \, k} \right) e^{-(M_Z/k)x} -1 \right) \tilde{\chi}_{2i}(x).
\label{eq:coupl}
\eea
We solve them numerically, obtaining quantitative agreement with the earlier literature \cite{Hisano:2004ds,Cohen:2013ama}.\footnote{Thank you to T.~Slatyer for providing a detailed comparison over a range of data points}.  There are nonetheless a couple provisos to the analysis.  Firstly, Eq.~\ref{eq:coupl} requires us to input a WIMP velocity.  In accord with previous references, we have chosen $v=10^{-3}$.  Scanning over a range of velocities, we found insensitivity to the precise number as long as it was below $M_W/M_\chi$ and we were not  in the immediate vicinity of the one of the resonance peaks in Fig.~\ref{somm}.  However, since the annihilation rate is so large at these peaks, a WIMP mass at these values is sufficiently ruled out to make this detail beyond our scope.  Additionally, the formal boundary conditions demand $\tilde{\chi}_{2i}(X) =0$ and calculate the Sommerfeld enhancement from $\xi_i(X)$, with $X \rightarrow \infty$, but in practice we must take finite $X$.  For our $M_\chi$ range of interest, the longest decay length for the charged state is determined by the $\delta M$ plus Coulomb part of the potential, and is given by $x_{\rm dec.}=1/\sqrt{2 M_\chi \delta M/k^2 - 1}$.  We find that taking $X = \mo(10) x_{\rm dec.}$ leads to numerical stability.

Looking at the results of our Sommerfeld analysis in Fig.~\ref{somm}, we see the expected resonance structure, with values of $|\psi(0)|^2$ that exceed $10^4$.  
The Sommerfeld factors $|\psi_2(0)|^2$ and $|\psi_1(0)|^2$ are comparable throughout our range and maintain the modest hierarchy of the perturbative charged-state annihilation due to $f_+$ exceeding $f_-$ and its contribution having a larger numerical prefactor in Eq.~\ref{eq:diffrate}.

\section{Dark Matter Constraints and Conclusion}
\label{sec:conc}

\begin{figure}[ht!]
\centerline{\scalebox{0.6}{\includegraphics{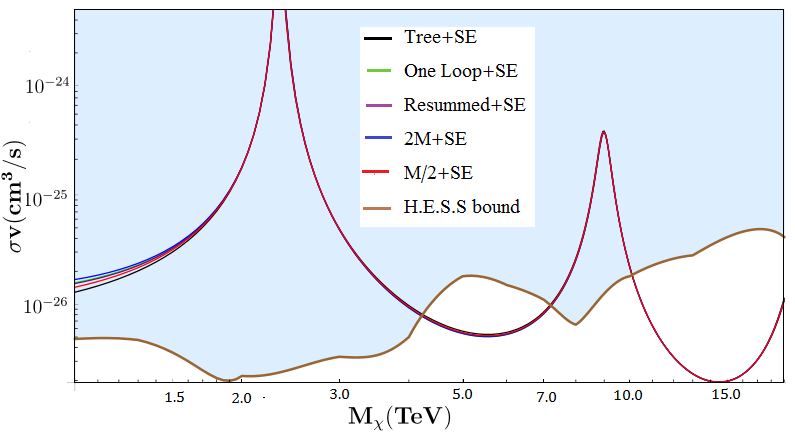}}}
\caption[1]{Annihilation cross section to $\gamma + X$. Exclusion taken from \cite{Ovanesyan:2014fwa}, assuming an NFW profile.} 
\label{sigma}
\end{figure}

Having calculated tree level matching, LL resummation, and computed the Sommerfeld enhancement numerically, we can now evaluate the differential cross section for $\chi^0 \chi^0 \rightarrow \gamma + X$, given in Eq.~\ref{eq:diffrate}.  We plot this in Fig.~\ref{sigma}, where we have digitized the HESS limits given \cite{Ovanesyan:2014fwa}.  
We note that in contrast to those groups that performed an exclusive two-body calculation, \cite{Bauer:2014ula,Ovanesyan:2014fwa}, we find the effect of higher order correction to be very modest.  For example, at the thermal relic mass of 3 TeV, we find
\bea
\langle \sigma v\rangle_{\rm LO}  &=& 5.4\times 10^{-26} \; {\rm cm^3/s}  \nn \\
\langle \sigma v\rangle_{\rm NLO-fixed}  &=& 5.3\times 10^{-26} \; {\rm cm^3/s} \nn \\  
\langle \sigma v\rangle_{\rm LL}  &=& 5.3\times 10^{-26} \; {\rm cm^3/s} , 
\eea
where for each value we have included Sommerfeld enhancement, and ``NLO-fixed'' includes only those one-loop effects that are resummed by our LL operator running.  
Thus, at this value for $M_\chi$, the leading corrections shift the semi-inclusive annihilation by just a few percent.  Comparing directly to \cite{Ovanesyan:2014fwa}, which also investigated annihilation of a triplet fermion, they find that at 3 TeV, higher order effects lead to a $\sim$50\% reduction in the exclusive rate.  This difference is to be expected given the distinct difference in our choice of observables.  From Eq.~\ref{eq:diffrate} and Fig.~\ref{somm}, we see that the leading contribution by a factor of few to our rate in our range of interest comes from perturbative $\chi^+ \chi^-$ annihilation and is proportional to $|\psi_{\pm}|^2$ and therefore $f_+ \big(= 1 + \exp(- \frac{3\alpha_W}{\pi}\log^2[\frac{M_W}{E_\gamma})]\big)$. Thus $f_+$, which has a tree level value of 2, drops to 1 in the limit $M_\chi,E_\gamma \rightarrow \infty$, meaning the $|\psi_{\pm}|^2$ term decreases by 50\% only in this {\it infinite} limit.  Furthermore, in this asymptotic case, $f_- \rightarrow 1$ from its tree-level value of 0, so the $|\psi_{00}|^2$ and $\psi_\pm \psi_{00}$ terms can boost overall rate above this 50\% reduction.  Comparing to the LL results in Eqs.~13-16 of \cite{Ovanesyan:2014fwa}, their Sudakov factor goes like $\exp(-\frac{\alpha_W}{\pi} \log^2(\frac{M_W}{E_\gamma}))$.  Thus, in the limit of infinite DM mass, their rate drops to 0.  This is expected from the general result that exclusive rates vanish in the limit of infinite energy.  

\begin{figure}[ht!]
\centerline{\scalebox{0.6}{\includegraphics{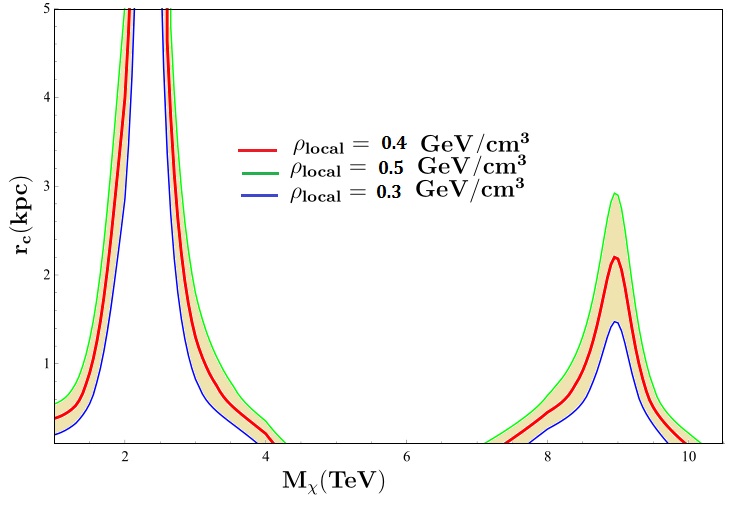}}}
\vskip-0.3cm
\caption[1]{The amount of coring required for the wino to become viable with respect to the HESS constraint shown in Fig.~\ref{sigma} for the cutoff-NFW profile (Eq.~\ref{eq:nfwcut}). The three curves display the effect of variation in the local dark matter density.} 
\label{coring}
\end{figure}

The limit from HESS in Fig.~\ref{sigma} shows that the thermal relic wino, $M_\chi \approx$ 3 TeV is ruled out by more than an order of magnitude.  Unfortunately, the astrophysical uncertainties in the halo profile are sufficient to evade an excess of even this size.  This is because the flux of photons measured by the  HESS  experiment is proportional to the ``$J$-factor'',
\beq
J = \frac 1 R_\odot \left( \frac 1 \rho_{\rm loc} \right)^2 \int_{\Delta \Omega} \int_{\rm l.o.s.} \rho^2(s,\Omega) ds,
\eeq
where $\rho_{\rm loc}$ is the local density, $R_\odot$ = 8.5 kpc is the distance to the galactic center, and $s$ is the line of sight distance to the experiment, where $r = \sqrt{s^2 + R_\odot^2 - 2 s R_\odot \cos\theta}$.  Discussions on the ability of different halo models to evade constraints can be found in the earlier papers that found the wino to be in tension with HESS \cite{Cohen:2013ama,Fan:2013faa}.  The exclusion curve we have taken from \cite{Ovanesyan:2014fwa} assumes an NFW profile \cite{Navarro:1995iw} with a local density, $\rho_{\rm loc} = 0.4$ GeV/cm$^3$ \cite{Catena:2009mf,Salucci:2010qr}, and $r_s$ = 20 kpc \cite{Iocco:2011jz},
\beq
\rho_{\rm NFW}(r) = \frac{\rho_0}{(r/r_s)(1+r/r_s)^2}.
\eeq
This is a cusped profile, diverging as $1/r$ toward the galactic center.\footnote{Typically, there will be deviations from strict spherical symmetry in the halo (axial, triaxial), and these can effect $J$ at the 10-20\% level \cite{Bernal:2014mmt}.  This uncertainty though, is far below the orders of magnitude shift in $J$ one can obtain by changing the profile shape between cusped and a large core.}  In the discussion that follows, we fix the local density $\rho_{\rm loc} = 0.4$ GeV/cm$^3$, but we will change the functional form of the distribution along with a possible core radius.  It is possible though, that the local density could lie somewhere in the range of 0.2-0.6 GeV/cm$^3$ \cite{Iocco:2011jz}.
The cusped vs.~cored (where the distribution flattens out at some distance) debate on the nature of the DM halo is an old one that experiment is far from resolving.  A recent observational analysis found good fits both for NFW profiles similar to the one in our constraint and for ones with relatively large, $\sim$10 kpc cores \cite{Nesti:2013uwa}.  Looking at simulation, DM-only models generically yield cusped distributions \cite{Pieri:2009je}.  However, observations of dwarf spheroidal galaxies found evidence for cored profiles, which was subsequently found in numerical models that included effects from baryons \cite{DiCintio:2013qxa}.  It is thought that supernovas near the galactic center may eject enough mass to flatten the DM distributions.  The question is whether sufficient mass could be ejected in a much more massive, $(100-1000)\times$ larger, Milky Way-like galaxy, or if the larger baryon density near the galactic center results in an even more cusped distribution.  Simulations of Milky Way-like galaxies that include baryons continue to show a preference for cusped distributions, at least into distances $\sim$1 kpc \cite{DiCintio:2013qxa,Kuhlen:2012qw,Marinacci:2013mha}. One can ask, how much coring is needed to save the wino, given our LL-resummed annihilation rate?  For an NFW profile that becomes constant below a certain radius,
\beq
\rho_{\rm cutoff-NFW}(r) = \left \{ \begin{array}{cc} 
\frac{\rho_0}{(r/r_s)(1+r/r_s)^2} & r>r_c \nn \\
\frac{\rho_0}{(r_c/r_s)(1+r_c/r_s)^2} & r\leq r_c 
\end{array}
\right.
\label{eq:nfwcut}
\eeq
in Fig.~\ref{coring} we plot the value of the core radius, $r_c$, needed to make our semi-inclusive rate calculation consistent with the limit from HESS.
For the thermal relic wino mass, $M_\chi$ = 3 TeV, if we consider the Burkert profile \cite{Burkert:1995yz,Salucci:2000ps},
\beq
\rho_{\rm Burk.}(r) = \frac{\rho_0}{(1+r/r_c)(1+(r/r_c)^2)},
\eeq
where $r_c$ again gives the core radius, and a cutoff-NFW distribution, we find that we need cores of 4-4.5 kpc and 1-1.5 kpc, respectively, to avoid the bounds.  Both distributions are well within the region allowed by observation \cite{Nesti:2013uwa}, and the wino in a cutoff-NFW profile (but not Burkert) is consistent with simulation, as well.  

The observation of wino dark matter near the thermal relic mass of 3 TeV would point to the existence of a nontrivial amount of coring in the halo of the galaxy which would require an explanation.  Of course, there are other possible ways to evade the HESS constraints, even if the profile were nearly NFW.  There is the possibility that the lightest neutralino may not be a pure wino.  For example, a thermal relic higgsino is far from constrained, and thus admixtures between these states could certainly be allowed \cite{Fan:2013faa}.  Sticking with the pure wino, if there were some non-thermal mechanism for its production, then the limit at values other than 3 TeV would be relevant, and $M_\chi$ could be in one of the allowed regions shown in Fig.~\ref{sigma}.  
\begin{figure}[ht!]
\centerline{\scalebox{0.7}{\includegraphics{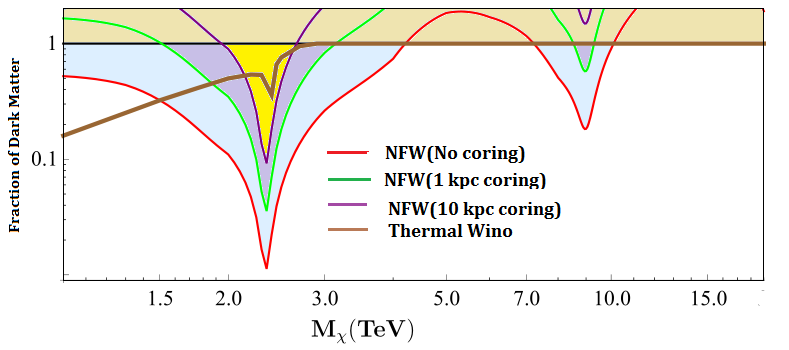}}}
\vskip-0.3cm
\caption[1]{Exclusion plot for an NFW profile with the wino making up only some fraction of the dark matter.  Expression for NFW profile with coring given in Eq.~\ref{eq:nfwcut}.} 
\label{fraction}
\end{figure}

Alternatively, whether or not its production were thermal, the wino could make up just a fraction of the dark matter, and thus much of parameter space would remain open, as shown in Fig.~\ref{fraction}.  With the theoretical uncertainty on its annihilation rate now under control at the $\mo(1\%)$ level,\footnote{It would be an interesting exercise to extend this analysis to NLL.  We have computed the running of our Wilson coefficients from the one-loop cusp anomalous dimensions.  One would also need one-loop non-cusp, two-loop cusp, and the $\beta$-function running of $\alpha_W$.  These were included in the exclusive-observable calculations of \cite{Bauer:2014ula,Ovanesyan:2014fwa}.  Additionally, the one-loop running of our fragmentation functions, Eq.~\ref{eq:fragdef}, is needed.} the discovery of a wino at future indirect detection experiments, such as CTA \cite{Doro:2012xx}, could give us important windows into further open questions such as the halo distribution, cosmological history of DM production, and the presence of multi-component dark matter.  

\vspace{0.1in}

\section*{Acknowledgements}
We thank Aneesh Manohar, Tracey Slatyer, Iain Stewart, and Brock Tweedie for discussions. The authors are supported by DOE grants DOE DE-FG02-04ER41338 and FG02-06ER41449.
M.B. acknowledges the Aspen Center for Physics where a portion of this work was completed.

\vspace{-0.2in}

\appendix
\section{Calculating the Schr\"odinger Potential from Quantum Field Theory}
\label{app:pot}

To compute the Sommerfeld enhancement, we solve the Schr\"odinger equation in the presence of the electroweak and mass shift potential,
\bea
V(r) = \left( \begin{array}{cc}
2\delta M -\frac{\alpha}{r} - \alpha_W c_W^2 \frac{ e^{-m_Z r}}{r} & -\sqrt{2}\alpha_W \frac{e^{-m_W r}}{r}   \\
-\sqrt{2}\alpha_W  \frac{e^{-m_W r}}{r}  &  0 
\end{array} \right).
\label{eq:potltwo}
\eea
This is an approximation to a full, quantum field theoretic treatment, but correctly sums all ladder-exchange diagrams \cite{Hisano:2004ds,Iengo:2009ni,Cassel:2009wt}.
To obtain Eq.~\ref{eq:potltwo} from the relativistic, electroweak lagrangian for an SU(2) triplet, we must define the two-body states evolved by the hamiltonian containing $V(r)$
in terms of the field theory creation and annihilation operators. After finding the four-fermion interaction in the nonrelativistic theory by matching to potential boson exchange, we can then find $V(r)$ by 
calculating how the interaction modifies our properly normalized, two-body states.  This allows us to write it in a form which makes no reference to fields, but evolves the state directly as a function of $r$.

We start with the standard lagrangian for a nonrelativistic field theory, which can be obtained for fermions by projecting to the single-particle sector of the Hilbert space (therefore decoupling the particle and antiparticle fields), pulling out a factor of $e^{-i M_\chi t}$
and then integrating out the small components of the fermion 4-spinor, leaving a field theory of 2-spinors,
\beq
\mathcal{L}_{\rm NRDM} = \chi^{0\dag} \left( iD_t + \frac{\bf{D}^2}{2M_\chi} \right) \chi^0 + \chi^{-\dag} \left( iD_t + \frac{\bf{D}^2}{2M_\chi} \right) \chi^- + \chi^{+\dag} \left( iD_t - \frac{\bf{D}^2}{2M_\chi} \right) \chi^+.
\label{eq:nrlag}
\eeq
We have adopted the convention that $\chi^0$ and $\chi^-$ in Eq.~\ref{eq:nrlag} destroy their respective particles, while $\chi^+$ is a creation field.  Explicitly, we have 
\bea
\chi^-(x) = \int \frac{d^3p}{(2\pi)^3} a_{p,s} \, e^{-ip\cdot x} \, \eta_s \nn \\
\chi^+(x) = \int \frac{d^3p}{(2\pi)^3} b^\dag_{p,s} \, e^{ip\cdot x} \, \eta_{-s} \nn \\
\chi^0(x) = \int \frac{d^3p}{(2\pi)^3} d_{p,s} \, e^{-ip\cdot x} \, \eta_s ,
\eea
where $p^0 = \mathbf{p}^2/2M_\chi$ in the Fourier factor and $\eta_s$ is a Pauli spinor with $\eta_1^\top = (1 \;\; 0),\,\eta_2^\top = (0 \;\; 1)$, and $\eta_{-1} = \eta_2,\,\eta_{-2} = \eta_1$.  Below, we use arrows to denote the spin.  Additionally, since $\chi^0$ is a Majorana fermion, it is useful to
define the following field, which also annihilates $\chi^0$ particle,
\beq
\chi^{0 \,c \; \dag}(x) = \int \frac{d^3p}{(2\pi)^3} d_{p,s} \, e^{-ip\cdot x} \, \eta_{-s}.
\eeq

\begin{figure}[ht!]
\centerline{\scalebox{1}{\includegraphics{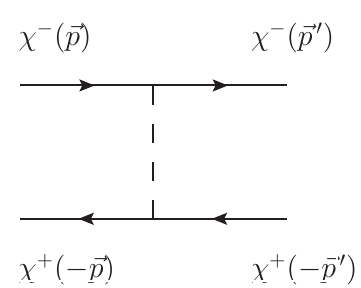}}}
\caption[1]{Amplitude for matching the exchange of a potential gauge boson in the full theory to a non-local four-fermion operator, Eq.~\ref{eq:qftpotl}} 
\label{potl}
\end{figure}

Matching to the full theory  we consider the exchange of a potential mode with propagator $i/(\mathbf{\Delta p}^2 + M^2)$, where $\mathbf{\Delta p = p^\prime-p}$ is the difference in spatial momentum of the 
incoming and outgoing fermions and $M$ is the mediator mass (Fig.~\ref{potl}).  
In terms of our nonrelativistic fields, and projecting to spin-singlets for our fermion bilinears via a Fierz transformation, we get
\bea
S_{\rm pot`l} &=& \int d^4x d^3y \, \frac{1}{2\,r} (\alpha + \alpha_W c_W^2 e^{-M_Z r}) [\chi^{-\dag}(x) \chi^+(x^0,\vec y)] [\chi^{+\dag}(x) \chi^-(x^0, \vec y)] \nn \\
&+& \frac{\alpha_W e^{-M_W r}}{2\,r} \big( [\chi^{0\dag}(x) \chi^{0\,c}(x^0, \vec y)] [\chi^{+\dag}(x) \chi^-(x^0,\vec y)] + {\rm h.c.}\big ),
\label{eq:qftpotl}
\eea
which agrees with \cite{Hisano:2004ds}.  To pass from nonrelativistic field theory to quantum mechanics, we need to construct our two-particle states.  Since the subtlety of obtaining Eq.~\ref{eq:potltwo} concerns the overall factors, we must pay special attention to the overall normalization.  For a state, $|\mathrm{B}(\vec p)_s \rangle$, of mass $M_B$ and spin $s$, we have that
\beq
\langle \mathrm{B}(\vec p^{\,\prime})_r |\mathrm{B}(\vec p)_s \rangle = 2 M_B (2\pi)^3 \delta^{(3)}(\vec p^{\,\prime} - \vec p) \delta_{rs}.
\label{eq:norm}
\eeq
Since we work in the CM frame of our two particles, there is no total momentum, and we therefore get the normalization factor $2 M_B (2\pi)^3 \delta^{(3)}(0) \delta_{rs}$.  Starting with our charged state, we satisfy Eq.~\ref{eq:norm} if we normalized it as
\beq
|\mathrm{C} \rangle = \sqrt{4 M_\chi} \int \frac{d^3 q}{(2\pi)^3} \psi_2(q) \frac{1}{\sqrt{2}} \left[a^\dag_{q,\uparrow} b^\dag_{-q,\downarrow} - a^\dag_{q,\downarrow} b^\dag_{-q,\uparrow} \right] |0 \rangle,
\eeq
where $\psi(q)$ is the relative momentum wavefunction, whose Fourier transform we will evolve with $V(r)$.  The calculation for the neutral, two-particle state is similar, but since $\chi^0$ is its own antiparticle, we get twice as many contractions, and the overall factor is therefore $\sqrt{2}$ smaller,
\beq
|\mathrm{N} \rangle = \sqrt{2 M_\chi} \int \frac{d^3 q}{(2\pi)^3} \psi_1(q) \frac{1}{\sqrt{2}} \left[d^\dag_{q,\uparrow} d^\dag_{-q,\downarrow} - d^\dag_{q,\downarrow} d^\dag_{-q,\uparrow} \right] |0 \rangle.
\eeq

Finally, extracting the field theory potential from $S_{\rm pot`l}$ in Eq.~\ref{eq:qftpotl}, we can check how the interacting hamiltonian acts on states $| \mathrm{N} \rangle, | \mathrm{C} \rangle$.  As an example, we take the off-diagonal interaction that converts a neutral state to a charged state and count the factors of 2.  The potential in $S_{\rm pot`l}$ contains an overall 1/2, and there are 4 possible contractions of the $d$ operators in the interaction with the $d^\dag$ operators in the state.  Finally the state $| \mathrm{N} \rangle$ has an overall normalization of $\sqrt{2M_\chi}$ and a $1/\sqrt{2}$ from the spin-singlet normalization.  Thus, we get 2$\times$ the creation operator structure of the $| \mathrm{C} \rangle$ state.  However, its normalization is an overall $\sqrt{4 M_\chi}$ times the $1/\sqrt{2}$ from being in the spin singlet, for an overall $\sqrt{2}$.  Thus, the potential that evolves $|\mathrm N \rangle$ into $|\mathrm C \rangle$ requires an overall $\sqrt{2}$, as we see in the ``12'' entry of Eq.~(\ref{eq:potltwo}).  Factors for the other entries follow similarly.  We see that the nontrivial $\sqrt{2}$ in the off-diagonal entry originates from the mismatch in normalization between the neutralinos, where the two-body state contains identical particles, and the chargino, where the two particles are distinct.

\end{document}